\documentclass[12pt,a4paper]{article}
\usepackage{amssymb,graphicx}
\newcommand{\ncpl}{{\mathbb T}^2_\frac{2}{N}}

\newcommand{\Tr}{\mbox{\rm Tr}}
\begin{document}

\title{Large-$N$ behavior of the IIB matrix model\\
       and the regularized Schild models}
\author{{\sc Naofumi Kitsunezaki}\thanks{Email address:
kitsune@eken.phys.nagoya-u.ac.jp}~~and
{\sc Shozo Uehara}\thanks{Email address:
uehara@eken.phys.nagoya-u.ac.jp}\\
{\it Department of Physics, Nagoya University,}\\
{\it Chikusa-ku, Nagoya 464-8602, Japan}}
\date{}
\maketitle
\vspace{-80mm}
\begin{flushright}
	DPNU-01-26\\
	hep-th/0108181\\
	August 2001
\end{flushright}
\vspace{50mm}

\begin{abstract}
We evaluate $N$ dependences of correlation functions in the
bosonic part of the IIB matrix model by the Monte Carlo method.
We also evaluate those in two sorts of regularized Schild models and
find that the $N$ dependences are different from those in the matrix
model.
In particular, the distribution of the eigenvalues are logarithmically
divergent in the regularized Schild model when $g^2N$ is fixed.
\end{abstract}

\section{Introduction}
The IIB matrix model \cite{IKKT} is the zero volume limit \cite{EK} of
a ten-dimensional super Yang-Mills theory, which is expected to be a
non-perturbative formulation of superstring theory.
The gauge fields in the IIB matrix model are represented by
$N\times N$ Hermitian matrices.
The distribution of eigenvalues of those matrices is interpreted
as our space-time.\footnote{This model represents an open space-time
since the eigenvalues of the matrices can take values in the infinite
region. Some studies of a closed space-time with periodic boundary
conditions were done with the unitary matrices instead \cite{KN}.}
Then, how the IIB matrix model settles the space-time structure, e.g.,
space-time dimension, is one of the most important problems.

We need to search for the true vacuum of the IIB matrix model to
solve the above problem.
When $N=2$, we can actually integrate the fermions and investigate
analytically the model with the bosonic degrees of freedom
\cite{TsuSu}. However, it is almost impossible to follow the same
procedure when $N>2$. So we should take other ways to analyze the
IIB matrix model.
It is natural to think that the IIB matrix model in the large-$N$
limit is described by a field theory because it has infinite numbers
of degrees of freedom.
We expect that we may be able to determine the true vacuum of the IIB
matrix model with the techniques of field theories if we could find
such a field theory.
We have paid attention to the fact that there is a one-to-one
correspondence between the action of an $N\times N$ matrix model and
that of the field theory on a non-commutative periodic lattice (NCPL)
with $N\times N$ sites, which we denote as $\ncpl$
\cite{bars,tokyou,landi,NBIs}.
Since the commutation relations of coordinates on NCPL are
proportional to $1/N$, one may think that the matrix model in the
large-$N$ limit has a one-to-one correspondence to a field theory
on the continuous torus.
Since the action of a naive large-$N$ limit of the IIB matrix model is
the supersymmetric Schild action on a torus \cite{Schild}, it is
believed that the finite-$N$ IIB matrix model is a regularized
supersymmetric Schild model \cite{deWitt}.
For this reason, the supersymmetric Schild model on a torus is a
plausible candidate for the large-$N$ limit of the IIB matrix model.

However, we pointed out a possibility that the large-$N$ behavior of
the IIB matrix model and that of the supersymmetric Schild model with
the momentum cut-off are different \cite{KU}.
And the differences between the bosonic matrix model and the
lattice-regularized Schild model were discussed in \cite{ANO}.
In our previous paper \cite{KU}, we proposed a method to estimate the
large-$N$ dependences of correlation functions of the IIB matrix model
kinematically. We also applied this method to the supersymmetric
Schild model with the momentum cut-off and found that it has a
different large-$N$ dependence from the IIB matrix model because its
action has a different overall factor of $N$.

In this paper, we confirm our conjecture by calculating correlation
functions using Monte Carlo simulation.
As for the bosonic matrix model, our Monte Carlo results are, of
course, consistent with those in Ref.\cite{HNT}.
We also calculate some correlation functions in two kinds of
regularized Schild model, one of which uses the lattice regularization
and the other introduces a naive momentum cut-off.
We find that the bosonic matrix model and both of the regularized
Schild model have different large-$N$ dependences.
Comparison between the large-$N$ dependences of the bosonic matrix
model and those of the lattice regularized Schild model were studied
before in Ref.\cite{ANO} and our results are consistent with theirs.
This paper is organized as follows. In the next section we show the
numerical results by the Monte Carlo simulation of the bosonic matrix
model and in section 3, we consider the regularized Schild models and
present the results by the Monte Carlo simulations.
The final section is devoted to summary and discussion.

\section{The bosonic matrix model}
In Ref.\cite{KU},  we proposed a method to estimate the large-$N$
dependences of correlation functions of the IIB matrix model
kinematically. Here we restrict our discussions on the bosonic part of
the IIB matrix model.
The action of the bosonic matrix model is given by
\begin{equation}
	S_{\rm M}=-\frac{1}{4g^2}\,\Tr\,[ A^\mu,A^\nu ]^2,
\end{equation}
where $A^\mu$ are $N\times N$ Hermitian matrices and $\mu,\nu =1,
\cdots,10.$\footnote{Partition function was evaluated to be finite
\cite{KNS,AW}.}
In this case our kinematical study leads to the result that any
correlation functions behave as
\begin{equation}
  \left\langle\frac{1}{N}\,\Tr\, F(A^\mu)\right\rangle
	\lesssim O(N^0),\label{eqn:conjecture}
\end{equation}
when $N$ goes to infinity keeping $g^2N$ fixed. This is the same
statement as 'tHooft argued perturbatively \cite{tHooft} and it was
shown perturbatively in Ref.\cite{HNT},  while we do not use any
perturbative arguments but only kinematical naive counting.
We should note that the naive counting does not lead to
eq.(\ref{eqn:conjecture}) with the usual matrix forms.
For example, when we calculate the $N$ dependence of
$\langle (1/N)\,\Tr\,(A^2)^2\rangle$ naively, we find
\begin{equation}
 \left\langle\frac{1}{N}\,\Tr\,(A^2)^2\right\rangle
  =\left\langle\frac{1}{N}\sum_{i,j,k,l=1}^N
    \left(A^\mu_{ij}A^\mu_{jk}A^\nu_{kl}A^\nu_{li}\right)
    \right\rangle\lesssim O(N^3),
\end{equation}
since the summation is over $N^4$ terms.
In general, we have
\begin{equation}
  \left\langle\frac{1}{N}\,\Tr\,F(A^\mu)\right\rangle
	\lesssim O(N^{p-1}),
\end{equation}
where $F(x)$ is a polynomial of degree $p$.
The situation is changed, however, when we map the IIB matrix model to
the field theory on NCPL \cite{bars,tokyou,landi,KU}.
We can rewrite any matrix models to field theories on NCPL
because the commutation relations of the plain waves on NCPL coincide
with U(N) algebra \cite{bars}. Correlation functions are rewritten by
\begin{equation}
  \left\langle\frac{1}{N}\,\Tr\,F(A^\mu)\right\rangle
	=\left\langle\frac{1}{N^2}
    \sum_{\sigma} F_\star (A^\mu(\sigma))\right\rangle,
    \label{eqn:NCPLcorfun}
\end{equation}
where $F_{*}(\cdot)$ has the same functional form as $F(\cdot)$ but the
products of the arguments are replaced by the star products on NCPL,
$\sigma$ are $N^2$ points on NCPL, and hence
eq.(\ref{eqn:NCPLcorfun}) is at most $O(N^0)$ because the summation
over $\sigma$ is the summation over $N^2$ points.

\begin{figure}[tb]
 \begin{center}
  \begin{tabular}{cc}
   \resizebox{78mm}{!}{\includegraphics{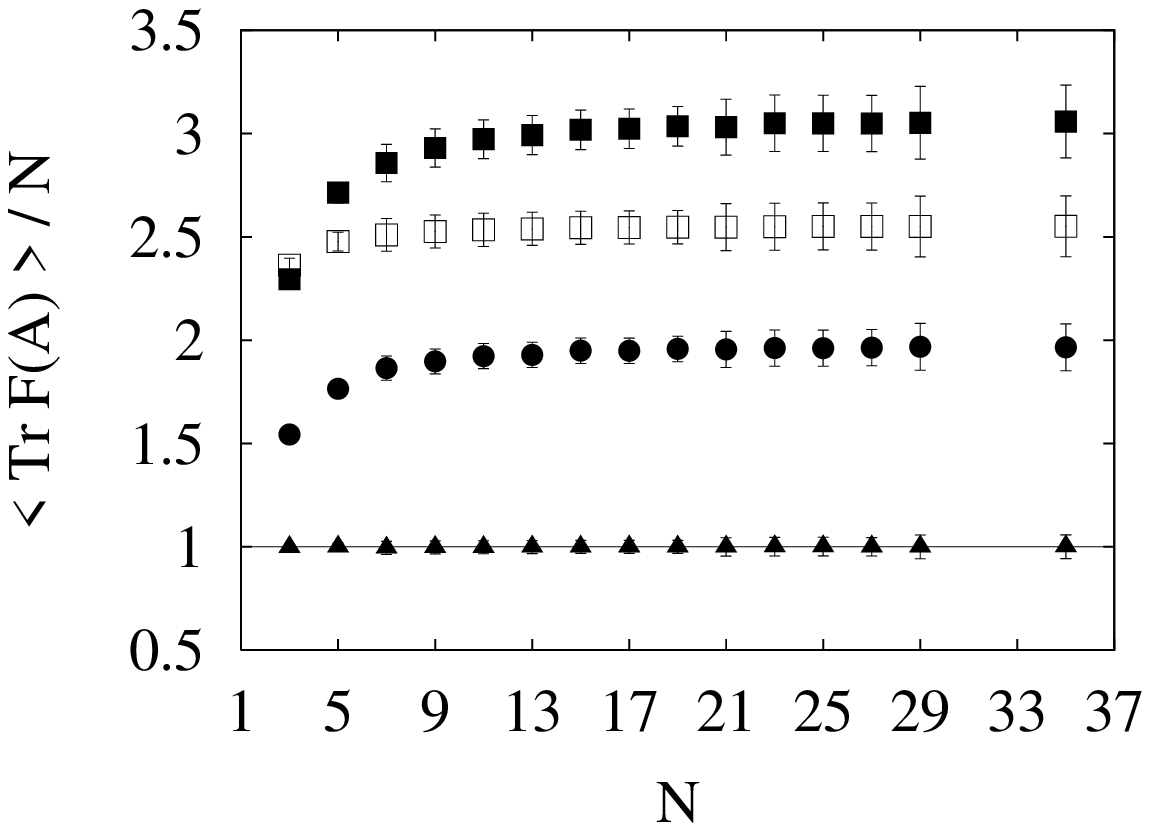}}&
   \resizebox{78mm}{!}{\includegraphics{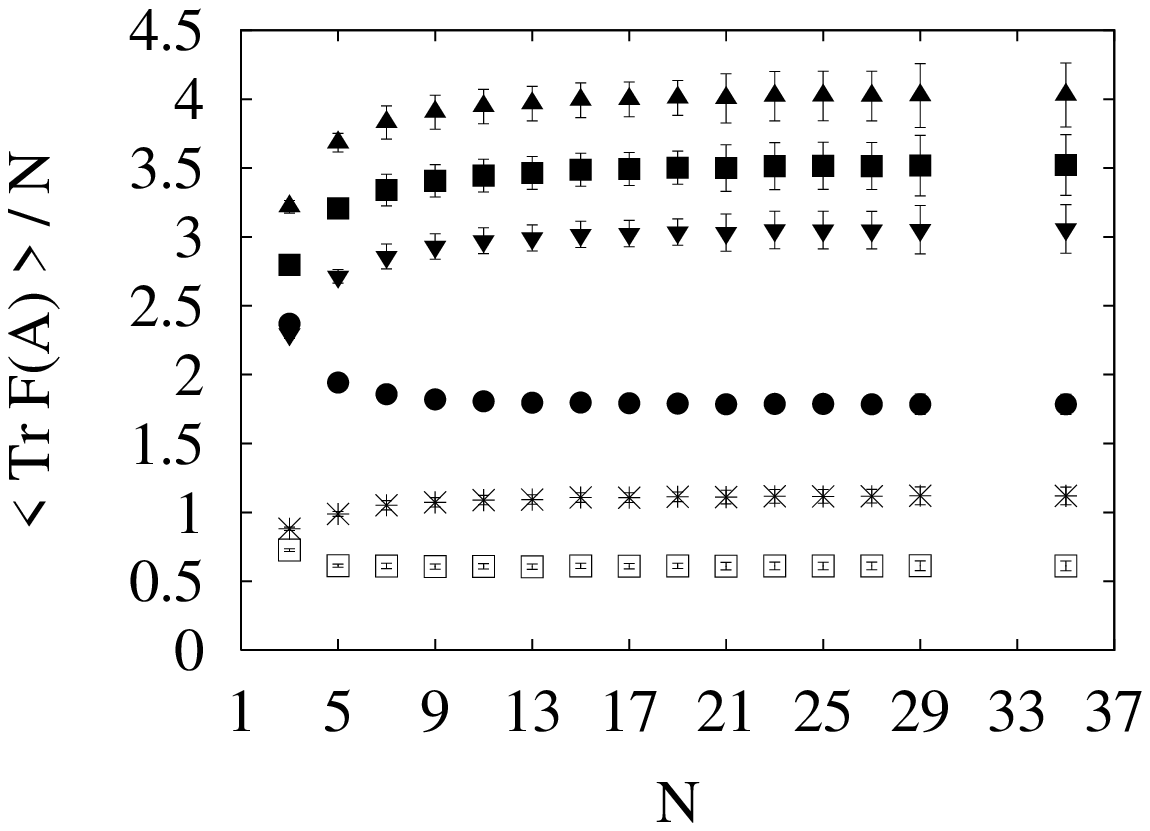}}
  \end{tabular}
\caption{The $N$ dependences of the correlation functions.
In the left figure, $\blacktriangle$ represents
$\langle\frac{1}{g^2} \Tr\,[A^\mu,A^\nu]^2\rangle /(10(N^2-1))$,
which can be analytically evaluated to be 1.
Other marks, $\blacksquare$, $\square$, $\bullet$, represent
$\langle\frac{1}{N}\,\Tr\,(A^2)^2\rangle /(g^2N)$,
$\langle\frac{1}{N}\,\Tr\,A^2\rangle/(gN^\frac{1}{2})$,
$\langle\frac{1}{N}\,\Tr\,(A^\mu A^\nu)^2\rangle/(g^2N)$,
respectively.
In the right figure, the marks, $\blacktriangle$, $\blacksquare$,
$\blacktriangledown$, $\bullet$, $*$, $\square$, represent
$\langle\frac{1}{N}\,\Tr\,(A^2)^3\rangle/(5g^3N^\frac{3}{2})$,
$\langle\frac{1}{N}\,\Tr\,(A^2A^\mu)^2\rangle/(5g^3N^\frac{3}{2})$,
$\langle\frac{1}{N}\,\Tr\,(A^2)^4\rangle /(20g^4N^2)$,
$\langle\frac{1}{N}\,\Tr\,(A^\mu A^\nu A^\rho)^2\rangle
	/(5g^3N^\frac{3}{2})$,
$\langle\frac{1}{N}\,\Tr\,A^2(A^\mu A^\nu)^2\rangle
	/(5g^3N^\frac{3}{2})$,
$\langle\frac{1}{N}\,\Tr\,(A^\mu A^\nu A^\rho A^\mu A^\rho
	A^\nu)\rangle /(5g^3N^\frac{3}{2})$,
respectively.}\label{fig:IIB}
\end{center}
\end{figure}

We calculate some correlation functions using Monte Carlo methods and
the results are shown in Fig.\ref{fig:IIB}.
The results in the left figure of Fig.\ref{fig:IIB} correspond to the
previous calculations in Ref.\cite{HNT}.
Since $\langle(1/g^2)\Tr [A^\mu,A^\nu]^2\rangle/(10(N^2-1))$ is
evaluated analytically to be 1, it was calculated to check on the
validity of our calculations and it gives perfectly the expected value
of 1.
The right figure of Fig.\ref{fig:IIB} shows the correlation
functions of all possible independent combinations of homogeneous
polynomials of degree six with rotational invariance and a particular
one of degree eight, $\langle(1/N)\,\Tr(A^2)^4\rangle$.
We calculate these correlation functions to show more evidences
for eq.(\ref{eqn:conjecture}).
When we fit those data to $a+bN^c$ by the least square method,
we find that $c$ is negative and $c=-2$ fits well \cite{HNT}.
Fig.\ref{fig:IIB} shows that all of the correlation functions
are convergent in the large-$N$ limit, which strongly supports
eq.(\ref{eqn:conjecture}).

\section{The regularized Schild models}
\begin{figure}[tb]
 \begin{center}
  \begin{tabular}{cc}
   \resizebox{78mm}{!}{\includegraphics{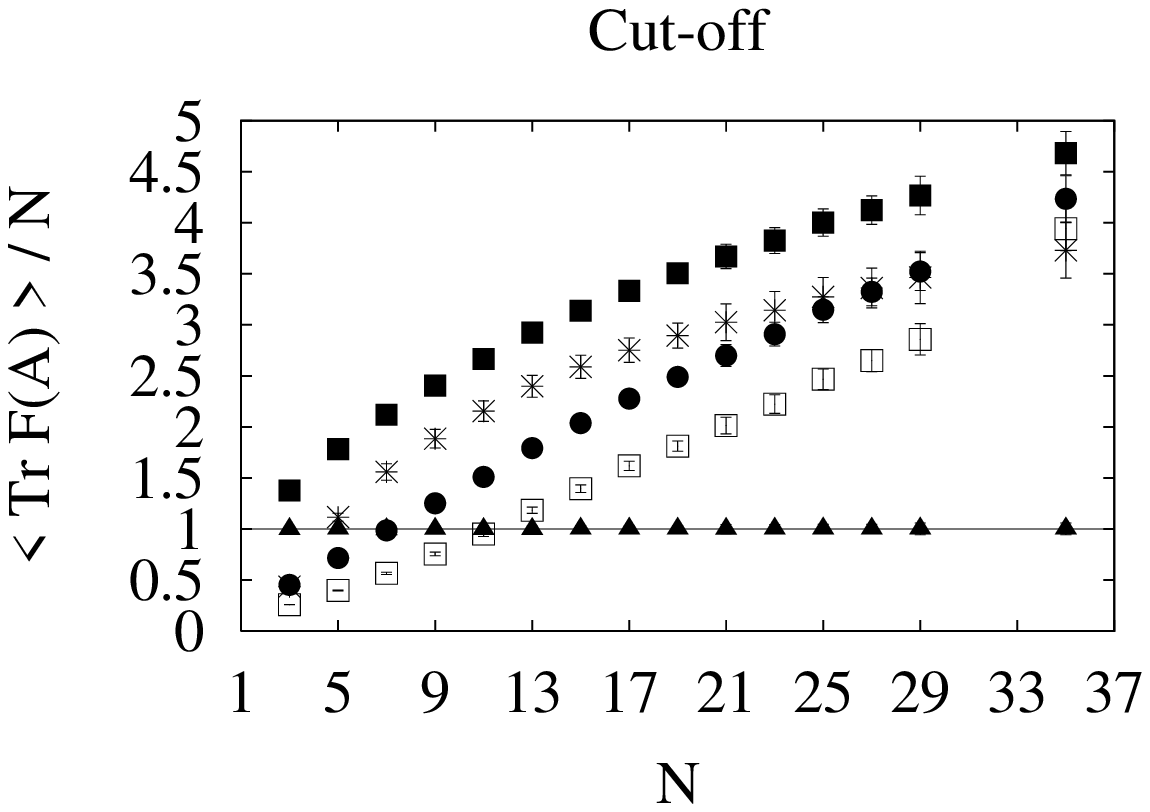}}&
   \resizebox{78mm}{!}{\includegraphics{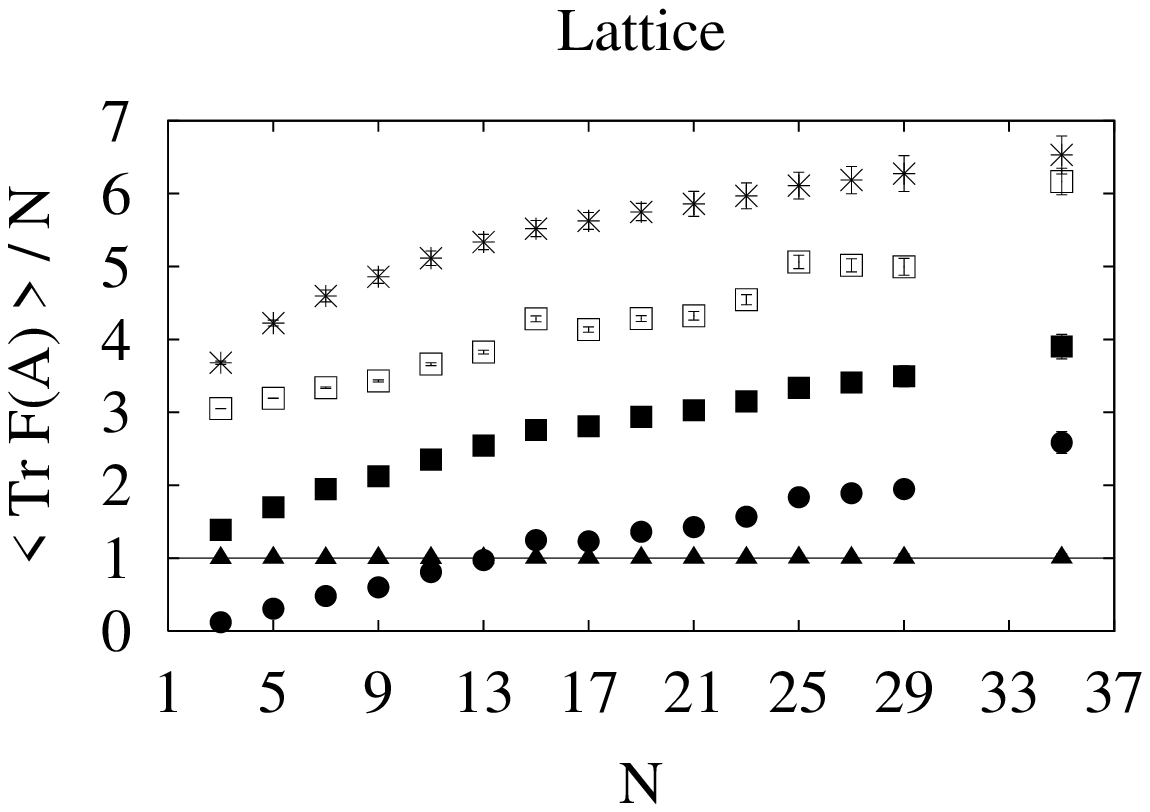}}
  \end{tabular}
 \caption{The $N$ dependences of correlation functions.
  The left figure is for the cut-off regularization, while the right
  one is for the lattice regularization. In both cases,
  $\blacktriangle$ represents
  $\langle\frac{1}{g^2N}\sum
  (\partial A^\mu(\sigma)\times\partial A^\nu(\sigma))^2\rangle
  /(10(N^2-1))$, which can be analytically evaluated to be 1.
  The marks, $*$, $\blacksquare$, $\bullet$, $\square$ represent
  $\langle\frac{1}{N^2} \sum   A^2(\sigma)\rangle/(gN^\frac{1}{2})$,
  $\langle\frac{1}{N^2} \sum (A^2(\sigma))^2\rangle/(g^2N)$,
  $\langle\frac{1}{N^2} \sum (A^2(\sigma))^3\rangle/(g^2N)$,
  $\langle\frac{1}{N^2} \sum (A^2(\sigma))^4\rangle/(g^2N)$,
  respectively.
  In fact each expectation value is divided by some constant numerical
  factor, respectively, so that all results can be seen in single
  frame at once.}\label{fig:Regs}
 \end{center}
\end{figure}

It is natural to think that the large-$N$ limit of the IIB matrix
model is the supersymmetric Schild model \cite{Schild,deWitt}.
This will be easy to understand when we rewrite the bosonic matrix
model to the field theoretical form on NCPL \cite{bars} as follows,
\begin{equation}
  S_{\rm M}=-\frac{1}{4g^2}\,\Tr\,[A^\mu,A^\nu]^2
    =\frac{1}{4g^2N}\sum_\sigma~[A^\mu(\sigma),
    A^\nu(\sigma)]_\star^2\,, \label{eqn:NCPLacction}
\end{equation}
where
\begin{eqnarray*}
  [A(\sigma),B(\sigma)]_\star&=&
	 A(\sigma)\star B(\sigma)-B(\sigma)\star A(\sigma),\\
  A(\sigma)\star B(\sigma)&=&A(\sigma)\,e^{-i\frac{1}{2\pi N}
	{\overleftarrow\partial}\times{\overrightarrow\partial}}
	 \,B(\sigma).
\end{eqnarray*}
In the large-$N$ limit, if we neglect naively the non-leading terms of
$1/N$, eq.(\ref{eqn:NCPLacction}) becomes
\begin{equation}
  \frac{1}{4\pi^2g^2N^3}\sum_\sigma
    \left(\partial A^\mu(\sigma)\times\partial A^\nu(\sigma)\right)^2
  \rightarrow \frac{1}{4\pi^2g^2N}\int_{T^2}d^2\sigma
    \left(\partial A^\mu(\sigma)\times\partial A^\nu(\sigma)\right)^2,
\end{equation}
and the last equation is exactly the Schild action on the torus.
The space where the fields live in is the two-dimensional torus because
the momenta of the plane waves in finite $N$ are discrete \cite{bars,KU}.
Thus the supersymmetric Schild model is a possible candidate for the
action of the IIB matrix model in the large-$N$ limit.
So we calculate correlation functions using some regularized
Schild model by Monte Carlo simulation to see whether they agree with
the results of the bosonic matrix model or not.
We adopt two kinds of regularizations for the Schild model, one is the
lattice regularization and the other is the momentum cut-off.
The field theoretical description of the finite-$N$ matrix model is
a field theory on NCPL with $N\times N$ sites, and the commutation
relation of the coordinates is proportional to $1/N$.
So it is natural to think that the finite but large-$N$ matrix
model approximately agrees with the Schild model regularized by the
commutative lattice with $N\times N$ sites.
On the other hand, the leading term of the field theoretical form of
the finite-$N$ matrix model action in momentum representation
exactly agrees with the Schild model action with the
momentum cut-off. Then we adopt these two regularizations to compare
with the finite-$N$ matrix model.
The actions of these regularized Schild model are as follows in the
momentum representation:
\begin{equation}
  S=\frac{N}{4g^2}\sum_{\mu,\nu=1}^{10}\sum_{\bf m,n,k}
    {\cal S}({\bf m-n},{\bf n}){\cal S}({\bf -m-k},{\bf k})\,
    A^\mu_{\bf m-n}A^\mu_{\bf -m-k}A^\nu_{\bf n}A^\nu_{\bf k},
\end{equation}
where {\bf m, n, k} are $(m_1,m_2)$, $(n_1,n_2)$, $(k_1,k_2)$,
respectively and run from $-(N-1)/2$ to $(N-1)/2$, and
\begin{eqnarray*}
  {\cal S}_{\mbox{\scriptsize cut-off}}({\bf m},{\bf n})
	 &=&\frac{2\pi}{N^2}\,{\bf m}\times{\bf n},\\
  {\cal S}_{\rm lattice}({\bf m},{\bf n})
	 &=&\frac{1}{2\pi}\,\epsilon^{ab}\sin\left(\frac{2\pi
	 m_a}{N}\right)\,\sin\left(\frac{2\pi n_b}{N}\right) ,
\end{eqnarray*}
where an extra $1/N$ factor has been multiplied in order to compare
with the matrix model clearly \cite{KU}.
Since the fields live in the commutative space here, $A^\mu(\sigma)
A^\nu(\sigma) = A^\nu(\sigma) A^\mu(\sigma)$, we calculate
$\langle(1/N^2)\sum (A^2(\sigma))^k\rangle /(g^2N)^{k/2}$,
$k=1,2,3,4$, to compare with the correlations functions of the bosonic
matrix model in Fig.\ref{fig:IIB}.
Also we calculate $\langle(1/g^2N)\sum (\partial A^\mu\times\partial
A^\nu)^2\rangle/(10(N^2-1))$ in order to check on the validity of our
calculations, since it is evaluated analytically to be 1.
The results of the correlation functions of these two regularized
models are given in Fig.\ref{fig:Regs}.
The correlation functions do not seem to be convergent so far, or it
looks divergent in the large-$N$ limit. In particular, when we fit the
data of the distribution of the eigenvalues to $a+bN^c$ by the least
square method, we find $c\sim 0$. Hence we put
\begin{equation}
  \frac{1}{gN^\frac{1}{2}}
  \left\langle\frac{1}{N^2}\sum_\sigma A^2(\sigma)\right\rangle
  \sim a\log(N)+b\,, \hspace{3ex}\mbox{($a,b$ : constants)}
  \label{eqn:log}
\end{equation}
and find that this fits well.
All these results are completely different from those of the bosonic
matrix model, and we can conclude that the bosonic matrix model is
{\it not} the regularization of the Schild model.
These results and conclusion are consistent with an early work
\cite{ANO}, where they compared the bosonic matrix model and the
lattice Schild model.

\section{Summary and Discussion}
We have calculated correlation functions of the bosonic matrix model
and two sorts of regularized Schild model by Monte Carlo methods to
investigate the large-$N$ limits of those models.
We have the following two conclusions:\\
(i) The Monte Carlo results of the bosonic matrix model strongly
supports eq.(\ref{eqn:conjecture}), which says that any correlation
functions such as $\langle (1/N)\,\Tr\,F(A)\rangle$ is at most
$O(N^0)$ when $N$ goes to infinity with $g^2N$ fixed to some
constant.\\
(ii) The $N$ dependences of correlation functions of the bosonic
matrix model do not agree with neither of those of regularized Schild
models.

Let us consider the first statement.
We see that the upper bound of the $N$ dependences of the
correlation functions can be calculated by naive counting.
In fact, there are some ways to count $N$ dependence naively and the
strongest constraint is eq.(\ref{eqn:conjecture}).
Let us demonstrate it with $\langle (1/N)\,\Tr\, (A^2)^2\rangle$ as an
example.
We have at least three different ways to express it by components of
matrices as follows:
\begin{eqnarray}
 \left\langle\frac{1}{N}\,\Tr\left((A^\mu)^2\right)^2\right\rangle
 &=&\left\langle\frac{1}{N}\sum_{i,j,k,l=1}^N
   \left(A^\mu_{ij}A^\mu_{jk}A^\nu_{kl}A^\nu_{li}\right)\right\rangle,
	\label{eqn:A4matrix}\\
 &=&\left\langle\frac{1}{N^2}\sum_{\sigma}\left(A^\mu(\sigma)^2\right)^2
	 \right\rangle,\label{eqn:A4NCPL}\\
 &=&\left\langle\sum_{{\bf m},{\bf n},{\bf k}}
   \left(A^\mu_{\bf m-n}A^\mu_{\bf n}A^\nu_{\bf -m-k}A^\nu_{\bf k}\right)
	 \right\rangle\label{eqn:A4momentum},
\end{eqnarray}
where ${\bf m}=(m_1,m_2), {\bf n}=(n_1,n_2), {\bf k}=(k_1,k_2).$
The r.h.s.\ of eq.(\ref{eqn:A4matrix}) is a usual matrix
representation, eq.(\ref{eqn:A4NCPL}) is the coordinate one on NCPL
and eq.(\ref{eqn:A4momentum}) is its momentum one.
The $N$ dependences by naive counting are at most $O(N^3)$, $O(N^0)$,
and $O(N^6)$, respectively.
In general, the upper bound of $N$ dependences of
$\langle(1/N)\,\Tr\,A^k\rangle$ by naive counting are given by
$O(N^k)$, $O(N^0)$, and $O(N^{2(k-1)})$ from the matrix
representation, the NCPL coordinate one, and its momentum one,
respectively.
The NCPL coordinate description, eq.(\ref{eqn:A4NCPL}), gives the
strongest constraint on the upper bound of large-$N$ dependence and
they agree with eq.(\ref{eqn:conjecture}).
We did not take a perturbative approach \cite{tHooft,HNT} but only
the kinematical one, and we expect that our arguments are
applicable to the supersymmetric one \cite{KU}.

Next let us discuss the second statement.
We should notice that the distributions of the eigenvalues,
$(g^2N)^{-1/2} \langle (1/N^2)\sum (A(\sigma))^2\rangle$,
diverge logarithmically in the regularized Schild models, while
it will be a constant in the large-$N$ limit of the bosonic matrix model.
In the IIB matrix model, $\langle (1/N^2)\sum (A(\sigma))^2\rangle$,
which is expected to be $O(g^2N)$, is interpreted as the extent of the
universe, while the fixed $g^2N$ will be proportional to $\alpha'^2$
\cite{FKKT}, so that the origin of the huge ratio should be explained.
On the other hand, if the extent of the universe diverges as in the
regularized Schild model, or we may fix $g^2N^{1-\kappa} \sim
\alpha'^2\  (\kappa>0)$ \cite{AIKKTT} in the bosonic matrix model,
there would be no problem of the huge ratio.
{}From this point of view, it is important to consider the possibility
that whether one can build the string theory from the supersymmetric
Schild model instead of the IIB matrix model as in Ref.\cite{FKKT}.


\end{document}